\documentclass[pdflatex,sn-mathphys]{sn-jnl}

\jyear{2022}%
\usepackage{xcolor}
\usepackage{anyfontsize}
\theoremstyle{thmstyleone}%
\usepackage{etoolbox}
\let\oldtextcolor\textcolor
\renewcommand{\textcolor}[2]{%
  \ifstrequal{#1}{red}%
    {#2}%
    {\oldtextcolor{#1}{#2}}%
}

\theoremstyle{thmstyletwo}%

\theoremstyle{thmstylethree}%

\begin{document}

\title[Neuroshaper]{Shaping freeform nanophotonic devices with geometric neural parameterization}

\author[1]{\fnm{Tianxiang} \sur{Dai}}

\author[1]{\fnm{Yixuan} \sur{Shao}}

\author[1]{\fnm{Chenkai} \sur{Mao}}

\author[1]{\fnm{Yu} \sur{Wu}}

\author[1]{\fnm{Sara} \sur{Azzouz}}

\author[2]{\fnm{You} \sur{Zhou}}

\author*[1]{\fnm{Jonathan A.} \sur{Fan}}\email{jonfan@stanford.edu}

\affil[1]{\orgdiv{Department of Electrical Engineering}, \orgname{Stanford University}, \city{Stanford}, \postcode{94305}, \state{CA}, \country{USA}}

\affil[2]{\orgdiv{Department of Physics and Optical Science}, \orgname{University of North Carolina at Charlotte}, \city{Charlotte}, \postcode{28223}, \state{NC}, \country{USA}}

\abstract{

Nanophotonic freeform design has the potential to push the performance of optical components to new limits, but there remains a challenge to effectively perform optimization while reliably enforcing design and manufacturing constraints.  We present Neuroshaper, a framework for freeform geometric parameterization in which nanophotonic device layouts are defined using an analytic neural network representation.  Neuroshaper serves as a qualitatively new way to perform shape optimization by capturing multi-scalar, freeform geometries in an overparameterized representation scheme, enabling effective optimization in a smoothened, high dimensional geometric design space.  We show that Neuroshaper can enforce constraints and topology manipulation in a manner where local constraints lead to global changes in device morphology. We further show numerically and experimentally that Neuroshaper can apply to a diversity of nanophotonic devices.  The versatility and capabilities of Neuroshaper reflect the ability of neural representation to augment concepts in topological design.

}

\maketitle

\section{Introduction}\label{intro}

Free space and on-chip nanophotonic systems have emerged as revolutionary platforms for controlling, routing, filtering, and transducing photons using subwavelength-scale structured media \cite{yu2014flat,kamali2018review,shrestha2018broadband,mohammadi2019inverse,guo2020squeeze,chen2021dielectric,hua2022ultra,zou2022pixel,zaidi2024metasurface,zhuochen2024ultracompact}.  The immense versatility of these platforms largely derives from the strong relationship between device geometry and optical response at subwavelength scales, which enables an exponentially large design space for optical manipulation.  Traditionally, these structure-function relations have derived from the use of simple, physically intuitive geometries in the form of nanoscale waveguides and resonators, together with their tailored coupling \cite{yu2011light,arbabi2015dielectric,koshelev2018asymmetric,rubin2019matrix,chen2019broadband,overvig2020multifunctional}.  Such simplifications have led to robust and scalable design rules for many classes of photonic systems, however, these concepts also pose fundamental limitations in performance and device footprint.  To overcome these limits, a wide range of freeform inverse design algorithms have been proposed that utilize a much larger design space of possible device geometries to achieve an objective \cite{SawyerMetaAtom,MingkunPerspective, gertler2023many,chen2024validation}.  The most advanced and computationally efficient algorithms utilize gradients calculated using the adjoint variables method or autodifferentiation to perform iterative optimization \cite{lalau2013adjoint,DavidSellLargeAngle,JianjiInitialLayout,JianjiMaterials,JianjiPhysics,chung2020high,shao2024multifunctional}.

A core feature of all inverse design methodologies is the specification of a geometric parameterization scheme that describes the device layout.  The choice of parameterization scheme is critical to reliably enforce constraints to the device geometry, in a manner that enables reliable device manufacturability and electromagnetic mode engineering control, and to ensure proper convergence of the optimization algorithm. Constraints of interest range from feature size and curvature tolerances to specification of the layout topology and connectivity.  There are currently three primary classes of parameterization schemes used in freeform inverse design.  One is pixel-based representations \cite{jensen2011topology, piggott2015inverse}, in which the device is described as an ultra-fine grid of pixels, each with a distinctly tailored dielectric constant value.  The second is level sets \cite{osher1988fronts, wang2003level},  in which layouts are implicitly represented as iso-contours of scalar functions \cite{wang2003level}.  The third is explicit geometric parameterization \cite{chen2020design, ErezReparam, zhou2024inverse}, in which simple analytic functions are defined to map latent variables to constrained geometric shapes.

While these parametrization concepts have been utilized effectively in certain design limits, each presents distinct trade-offs between optimization capability and constraints compliance.  Pixel-based methods are the most straightforward to adapt to gradient-based optimizers such as the adjoint variables method.  However, the use of spatial filters \cite{sigmund2007morphology}, density filters, and projection methods \cite{guest2004achieving} to impose constraints lead to either poor fabrication compliance or the imposition of overly restrictive constraints.  Traditional level set methods offer improved topology control, but they struggle with multi-scale features, require careful numerical schemes to maintain stability \cite{osher2001level}, and are limited in performance by their reliance on local shape derivatives \cite{allaire2004structural}. Explicit parameterization schemes have the benefit of hard coding strict design constraints in the parameterization functions themselves, but they are ultimately overly restrictive and suffer from limited topology control.  Qualitatively new geometric parameterization concepts are needed to  generally and robustly enforce multiple competing objectives and constraints in the device, while maintaining access to the large freeform design landscape.

In this work, we present Neuroshaper, a neural network-based approach to freeform geometric representation that can streamline with existing gradient-based freeform optimizers.  Neuroshaper is inspired by recent work in computational imaging and computer graphics on neural representations \cite{mildenhall2020nerf, park2019deepsdf} and specifically the development of multi-resolution techniques for representing complex signals \cite{muller2022instant}.  \textcolor{red}{Unlike traditional deep learning models that are trained on large datasets, Neuroshaper functions as a differentiable geometric parameterization scheme; its neural network parameters directly define the device geometry and are optimized using physics-based gradients and constraint penalties.}  It addresses limitations with existing parameterization schemes in photonics in multiple ways.  First, it takes advantage of the immense expressivity of neural networks paired with learnable coordinate encoding to capture a wide range of multi-scalar freeform shapes.  Second, it is differentiable and uses network loss function engineering and backpropagation to robustly and flexibly enforce constraints.  Third, by capturing geometric device parameters using large numbers of neural network weights, Neuroshaper serves as an overparameterized representation scheme that captures a smoothened, high-dimensional geometric design space.  In this manner, gradient-based optimizers are able to more easily traverse the design space and detour around forbidden zones when constraints are encountered (Figure \ref{fig:1}a). An example of this optimization trajectory is included in the Supplementary Movie 1. Fourth, Neuroshaper captures a global representation of geometric shapes, such that local geometric constraints map onto global modifications to the device layout.  With these collective advantages, devices designed using Neuroshaper exhibit better handling and enforcement of constraints, and also similar or better performance, compared to conventionally parameterized freeform devices.

\begin{figure}[htb]
\centering
\includegraphics[width=1.0\textwidth]{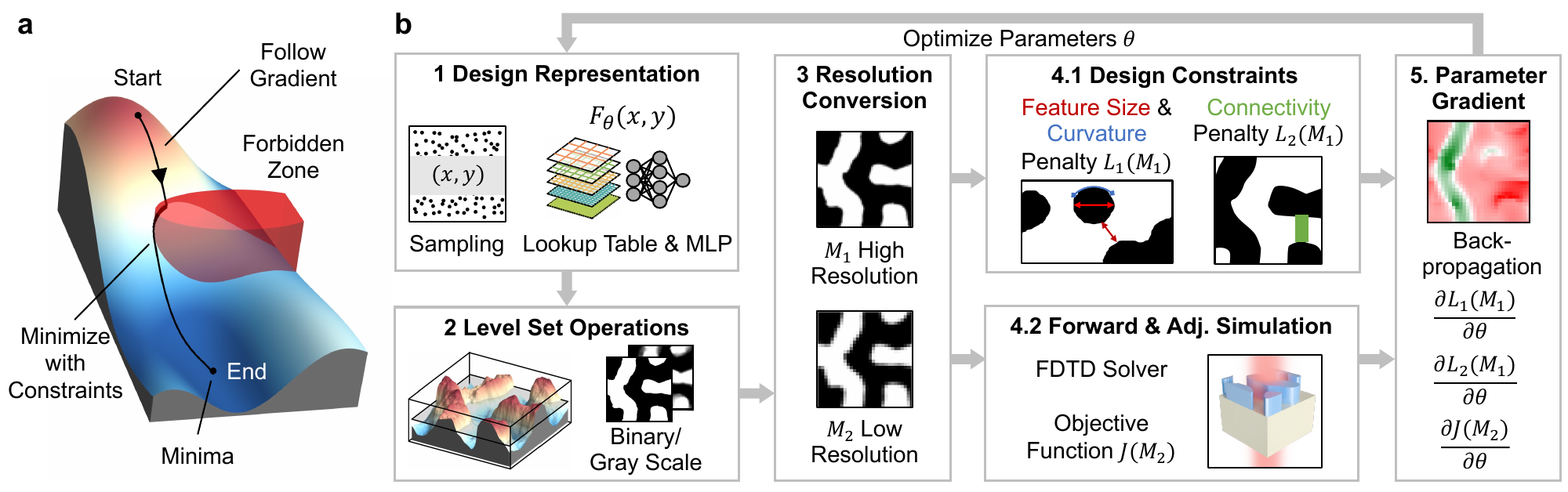}
\caption{Neuroshaper overview. a) Visualization of gradient-based optimization in the high dimensional Neuroshaper parameterization space. The optimizer avoids forbidden zones imposed by constraints to reach a suitable local minima. b) Overview of the Neuroshaper framework for nanophotonic device design. The pipeline consists of: (1) Neural design representation combining sampling points with lookup tables and multilayer perceptrons to generate continuous level set functions; (2) Level set operations to obtain binary/grayscale patterns; (3) Multi-resolution conversion between high ($M_1$) and low ($M_2$) resolutions; (4) Design evaluation based on geometric constraints and electromagnetic performance; (5) Parameter gradient computation via backpropagation.}
\label{fig:1}
\end{figure}

The full framework for Neuroshaper-enabled freeform inverse design is illustrated in Figure \ref{fig:1}b.  Freeform layouts are produced using a novel neural architecture that combines multi-resolution hash encoding with multilayer perceptrons to generate continuous level set functions. These functions undergo level set operations to produce binary or grayscale patterns suitable for both constraint evaluation and simulation. The ability for Neuroshaper to support multi-resolution representations enables high resolution representations to be used for geometric constraint evaluation ($M_1$) and low resolution versions for electromagnetic simulations ($M_2$). Global loss functions accounting for feature size, curvature, and connectivity constraints are aggregated together with electromagnetic performance to properly balance these different competing objectives.  All aspects of the algorithm are differentiable, enabling direct compatibility with autodifferentiation packages on GPU hardware and efficient gradient-based optimization of the complete system. \textcolor{red}{The 'electromagnetic performance' objective function ($J(M_2)$) is flexible and can be tailored to various goals, including maximizing transmission/conversion efficiency (as shown in Figure~\ref{fig:5}), minimizing reflection, or optimizing for phase-dependent characteristics (e.g., specific diffraction orders, wavefront shaping), provided the chosen simulation method allows for gradient computation with respect to the device geometry.}

\section{Results}\label{sec:results}
\subsection{Neuroshaper architecture}

\begin{figure}[htb]
\centering
\includegraphics[width=1.0\textwidth]{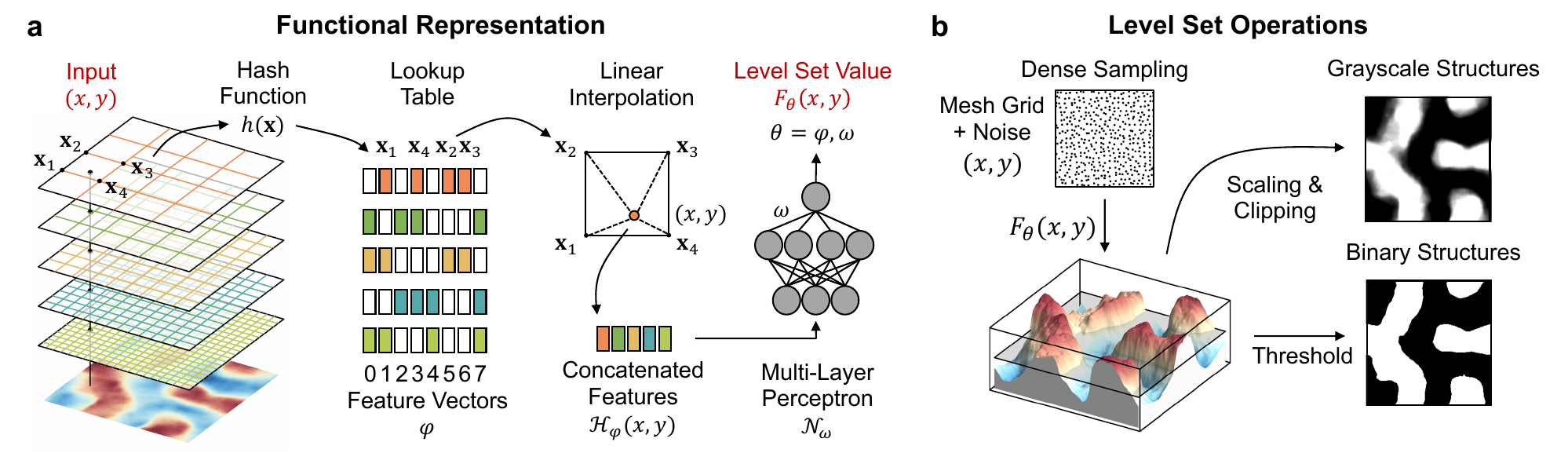}
\caption{The neural level set representation model. a) Functional representation showing the multi-resolution hash encoding process. Input coordinates $(x,y)$ are mapped through parallel hash functions at different resolution levels, each accessing a dedicated lookup table of trainable feature vectors. Features are interpolated based on the relative position within grid cells and concatenated. The resulting feature vector is processed by a multi-layer perceptron (MLP) to produce the final level set function value $F_\theta(x,y)$. b) Level set operations showing the conversion from continuous functions to binary and grayscale structures through dense sampling and thresholding operations.}
\label{fig:2}
\end{figure}

Schematics detailing the specification of nanophotonic geometries with Neuroshaper are presented in Figure~\ref{fig:2} and consist of two parts, level set encoding and level set translation to a dielectric distribution.  We focus on the encoding of level sets as an intermediary to dielectric layouts, as opposed to the direct encoding of dielectric layouts, due to the uniquely flexible and expressive nature of level sets in topology optimization. To specify level sets, we propose the use of a neural network-encoded hash table description that combines discrete feature vectors with continuous neural networks and allows for both local and global control of the design space while maintaining differentiability (Figure~\ref{fig:2}a). 

More specifically, we first find the nearest lattices of the input coordinates $(x,y)$ at every resolution level, represented as $\mathbf{x_{1,2,3,4}}$. In each level, the lattice point $\mathbf{x}$ is mapped to an entry in the lookup table by the hash function $h(\mathbf{x})$, which stores learnable feature vectors $\varphi$. These vectors from each hash table are then interpolated and concatenated to produce an encoded, high-dimensional feature vector, $\mathcal{H}_\varphi(x,y)$, which serves as an input to a multi-layer perceptron (MLP) that maps these features to scalar level sets.  The neural network mapping is described by the transformation $\mathcal{N}_\omega$, where $\omega$ are weights of the network.  Overall, the collective mapping of input coordinate to level set is specified by:
\begin{equation}
F_\theta(x,y) = \mathcal{N}_\omega(\mathcal{H}_\varphi(x,y))
\end{equation}
where $\theta = {\varphi, \omega}$ captures the hash table feature parameters and neural network parameters.  It is important to note that this neural network serves as a \textit{differentiable geometric representation} whose parameters $\theta$ are directly optimized based on simulation results and constraints, rather than a predictive model trained on a pre-existing dataset.  These parameters are learned using autodifferentiation and gradient-based updating with tailored loss functions, which will be discussed later.  Detailed mathematical derivations and implementation specifics are provided in Methods Section.

To translate the level set to a dielectric distribution (Figure~\ref{fig:2}b), differentiable operations are applied, and the Neuroshaper platform is capable of readily generating a range of grayscale and binary dielectric representations.  To produce a  grayscale dielectric structure layout, $M_g(x,y)$, scaling and clipping operations are used. Binary dielectric structures, $M_b(x,y)$, are generated using direct thresholding.

Multi-resolution hash encoding addresses fundamental limitations posed by traditional level set methods. The hash encoding effectively acts as an adaptive grid, automatically allocating more representational capacity to regions with important features through collision-based feature sharing.  As such, we achieve natural scale separation while maintaining differentiability. Our encoding concept also enables efficient parameter updates during optimization, as each input affects only a small subset of feature vectors and a small MLP network, but still produces both global and local changes in geometry due to the multiple resolution levels. In addition, the differentiability of our representation enables straightforward computation of gradients and curvature at each point.

\subsection{Initialization and Constraints}

Neuroshaper features a comprehensive system of methods for enforcing  constraints while maintaining optimization flexibility, and we provide an overview of these methods below.  The methods can be used in different combinations depending on the problem with no restrictions. Mathematical and implementation details are provided in Methods Section.

\begin{figure}[htb]
\centering
\includegraphics[width=1.0\textwidth]{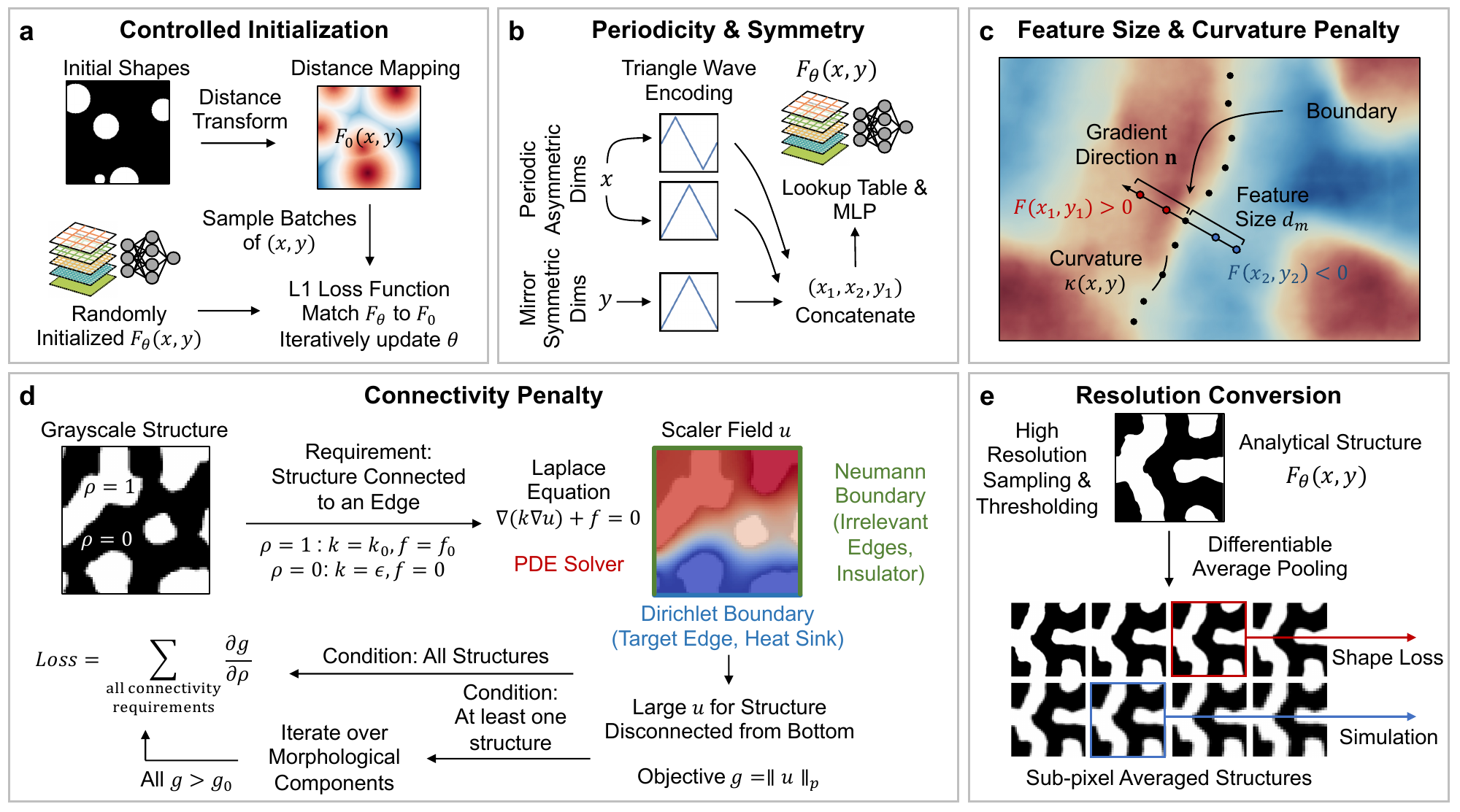}
\caption{Implementation of initialization methods, constraints and resolution conversions. a) Controlled initialization showing the transformation from initial shapes to distance mapping and the training process through randomly sampled batches. b) Implementation of periodicity and symmetry constraints using triangle wave functions and coordinate mapping. c) Feature size and curvature penalty calculation demonstrating gradient-based measurement of local geometric properties. d) Connectivity penalty implementation using PDE-based approach with specified boundary conditions. e) Resolution conversion process illustrating the relationship between high-resolution reference structures and differentiable sub-pixel averaged structures at different resolution for simulation.}
\label{fig:3}
\end{figure}

\textit{Initialization.} To initialize Neuroshaper, the parameters $\theta$ are randomly initialized from a normal distribution, and the corresponding level can serve as a random starting point for grayscale topology optimization. 
Neuroshaper can also be initialized to a desired shape using a two-stage process (Figure~\ref{fig:3}a). Here, the level set for the desired initial shape is first generated as an approximate level set function using a distance transform. The Neuroshaper parameters are then optimized using an L1 loss function to approximate this initial level set. This approach provides controlled initialization while also introducing slight randomness that breaks the symmetry of the level set and associate gradient calculations.

\textit{Periodicity and reflection symmetry.} Neuroshaper is capable of explicitly enforcing periodicity and reflection symmetry constraints by the implementation of a triangle wave coordinate mapping of the $x$ and $y$ coordinates prior to hash table encoding (Figure~\ref{fig:3}b). Triangle waves are specifically chosen for their constant gradient magnitudes, which preserve the multi-resolution characteristics of our design representation. This is in contrast with sinusoidal mappings, whose varying gradients can distort the multi-resolution nature of the representation.  Different types of triangle waves can be judiciously chosen to specify the reflection symmetry axes or to specify periodicity without any reflection symmetry.

\textit{Feature size and curvature constraints.} Feature size and curvature control are enforced through gradient-based boundary analysis (Figure~\ref{fig:3}c). Given a minimum feature size $d_m$ and a sampled boundary point $\mathbf{p}$ with a normal vector $\mathbf{n}$ into the void region, a loss function is defined such that all points from $\mathbf{p}$ to
$\mathbf{p}_+ = \mathbf{p} + d_m\mathbf{n}$ are specified to have positive values and those from $\mathbf{p}$ to
$\mathbf{p}_- = \mathbf{p} - d_m\mathbf{n}$ have negative values.
We also require that the level set is similar to a signed distance function of the boundary, which guarantees that after level set thresholding, all physical features in both material and void regions do not violate our minimum feature size constraint and are minimally modified each iteration. For curvature control, we leverage the differentiability of our shape representation to compute local curvature and add a loss function penalty term to discourage small values for local radii.

\textit{Connectivity.} Connectivity control is achieved by formulating connectivity as an auxiliary heat conduction problem (Figure~\ref{fig:3}d) in which regions requiring connectivity are treated as heat sources while the target connection points act as heat sinks \cite{li2016structural}. The steady-state heat equation defines the temperature distribution:
\begin{equation}
-\nabla \cdot (k\nabla u) = f
\end{equation}
where $u$ represents the temperature field, $k$ is the thermal conductivity determined by the binary design, and $f$ describes heat sources and sinks placed at desired connection points. 
Upon solving the heat equation, the resulting temperature distribution identifies disconnected regions through high temperature values, and the temperature gradient represents optimal paths for establishing connectivity and is utilized in the connectivity constraint loss function.  The connectivity constraint can be used to control device topology, be specified to global or local features, and can be applied to both material and void regions.

\textit{Multi-resolution scaling.} Neuroshaper can produce device layouts with different spatial resolutions (Fig.~\ref{fig:3}e), which is useful when considering that geometric constraint evaluation demands high-resolution structures ($M_1$) for accurate feature analysis while electromagnetic simulations can be efficiently and accurately performed at lower spatial resolutions ($M_2$).

To derive device layouts with different resolutions from the continuous function $F_\theta(x,y)$, we first sample this function at an extremely high base resolution $M_b$ that far exceeds both the geometric analysis and simulation requirements. From this high-fidelity binary representation, we derive both the geometric analysis resolution $M_1$ and the simulation resolution $M_2$ through average pooling operations with different downsampling factors. This enables sub-pixel averaging, which is crucial for maintaining the accuracy of electromagnetic simulations as it provides effective material properties that are consistent with the finite difference formalism and that well approximate the structural layout.

Our approach additionally overcomes issues pertaining to the introduction of aliasing artifacts upon changes in resolution. Since our representation remains analytically continuous throughout the optimization process, resolution changes merely affect the sampling density rather than the underlying structure. The difference becomes particularly evident in designs requiring highly precise geometric features.

\subsection{Demonstrations}

\begin{figure}[b]
\centering
\includegraphics[width=1.0\textwidth]{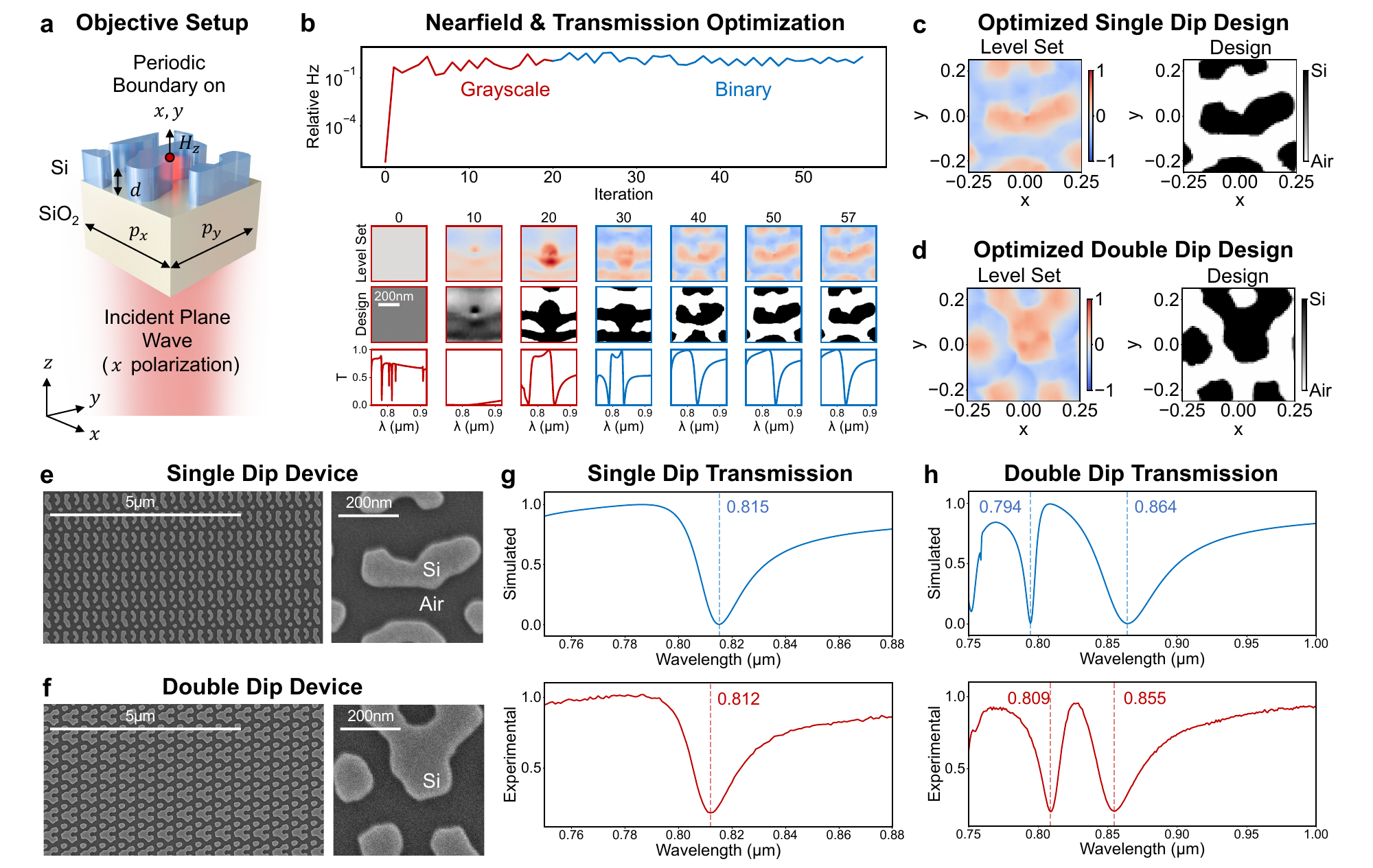}
\caption{Optimization setup and experimental validation for Neuroshaper-designed nonlocal metasurfaces. a) Schematic of the simulation setup showing a periodic silicon (Si) metasurface of thickness $d$ on a silica ($\text{SiO}_2$) substrate.  b) Evolution of the optimization process showing grayscale (red) and binary (blue) design regimes. The lower panels show the level set function (top), corresponding physical designs (middle), and transmission spectra (bottom) at different iterations. c,d) Level set function and binary silicon structure for the optimized (c) single-dip and double-dip design.  e,f) SEM images of the fabricated (e) single-dip and (f) double-dip devices.  g,h) Simulated (top) and experimental (bottom) transmission spectra for the (g) single-dip and (h) double-dip devices.  }
\label{fig:4}
\end{figure}

To demonstrate the different features of Neuroshaper and its ability to produce manufacturable nanophotonic devices, we design and experimentally implement freeform nonlocal metasurfaces serving as narrowband wavelength filters.  Nonlocal metasurfaces are devices that support the coupling of free space light into guided mode resonances \cite{zhou2023multi, kwon2018nonlocal, shastri2023nonlocal}, enabling light-matter interactions that are inherently sensitive to wavelength and incidence angle. We focus our demonstration on near-infrared periodic silicon-based metasurfaces that feature zeroth-order nonlocal responses.  To perform this optimization, we consider a modal framework for nonlocal optical phenomena in which the strong coupling of incident light to a vertically oriented magnetic dipole mode within a metasurface unit cell leads to a nonlocal response \cite{zhou2024freeform}.  To explicitly enforce this mode behavior, we perform near-field optimization to maximize the magnitude of the $H_z$ field component at a single point just above the metasurface, to avoid diverging values when performing adjoint gradient computation (Figure~\ref{fig:4}a).  For devices supporting multi-wavelength responses, multi-objective optimization is performed to specify modes at different wavelengths.  

Details pertaining to our optimization implementation can be found in Methods Section.  To summarize, we consider periodic devices with a unit cell period of 500 nm and normally incident light with linear polarization along the x-axis.  The adjoint variables method is used to perform near-field freeform optimization, and Tidy3D \cite{tidy3d} is used as the solver.  The iterative optimization process contains two parts (Figure~\ref{fig:4}b).  First, grayscale optimization is performed in which the continuous level set function is used to specify device layouts with grayscale dielectric values.  This device parameterization enables a thorough exploration of the design space and gradual convergence toward promising topologies. Second, iterative optimization of layouts with binary dielectric values is performed to fine-tune these devices into final configuration. 

Our optimization pipeline utilizes many of the constraints and topology control features in Neuroshaper to ensure the final devices support clean nonlocal behavior.  Triangle wave encoding is used to enforce periodic boundary conditions along both in-plane axes.  Minimum feature size and curvature constraints are implemented to limit the feature size and radii of curvature to 100 nm and 50 nm, respectively, which ensures that the device can be reliably fabricated.  With the connectivity constraint, we specify the device topology to be isolated silicon island structures, which helps to suppress the presence of laterally distributed optical modes that can spoil the background.  We note that our method for imposing this constraint is less restrictive than simply specifying the unit cell boundaries to be void, which excludes island structures that span across unit cell boundaries.  We additionally specify the device to be only 120 nm thick, which is much smaller than the target wavelength and which suppresses the presence of higher order modes spanning the device height. The small thickness also enhances manufacturing feasibility.

Single and double-dip metasurface designs are shown in Figures~\ref{fig:4}c,d.  
To fabricate these devices, we deposit a thin film of silicon onto a glass substrate by chemical vapor deposition, pattern a hydrogen silsesquioxane mask on the film using electron beam lithography, and transfer the pattern into the film using reactive ion etching.  Electron microscopy images of both devices (Fig.\ref{fig:4}e,f) reveal excellent fabrication fidelity and accurate matching with the design layouts.  To characterize the devices, transmission spectra are obtained by polarizing and illuminating each device under normal incidence using a tuneable laser and scanning wavelength in 1 nm increments. The measured transmission spectra (Fig.\ref{fig:4}g,h) show good agreement with simulations.  The single dip device shows a sharp transmission dip near its design wavelength of $\lambda$ = 815 nm and a clean background, while the double-dip device shows sharp transmission dips close to its design wavelengths of 800 nm and 860 nm. Additional details pertaining to the fabrication and characterization steps can be found in Section \ref{sec:setup}.

\begin{figure}[htb]
\centering
\includegraphics[width=1.0\textwidth]{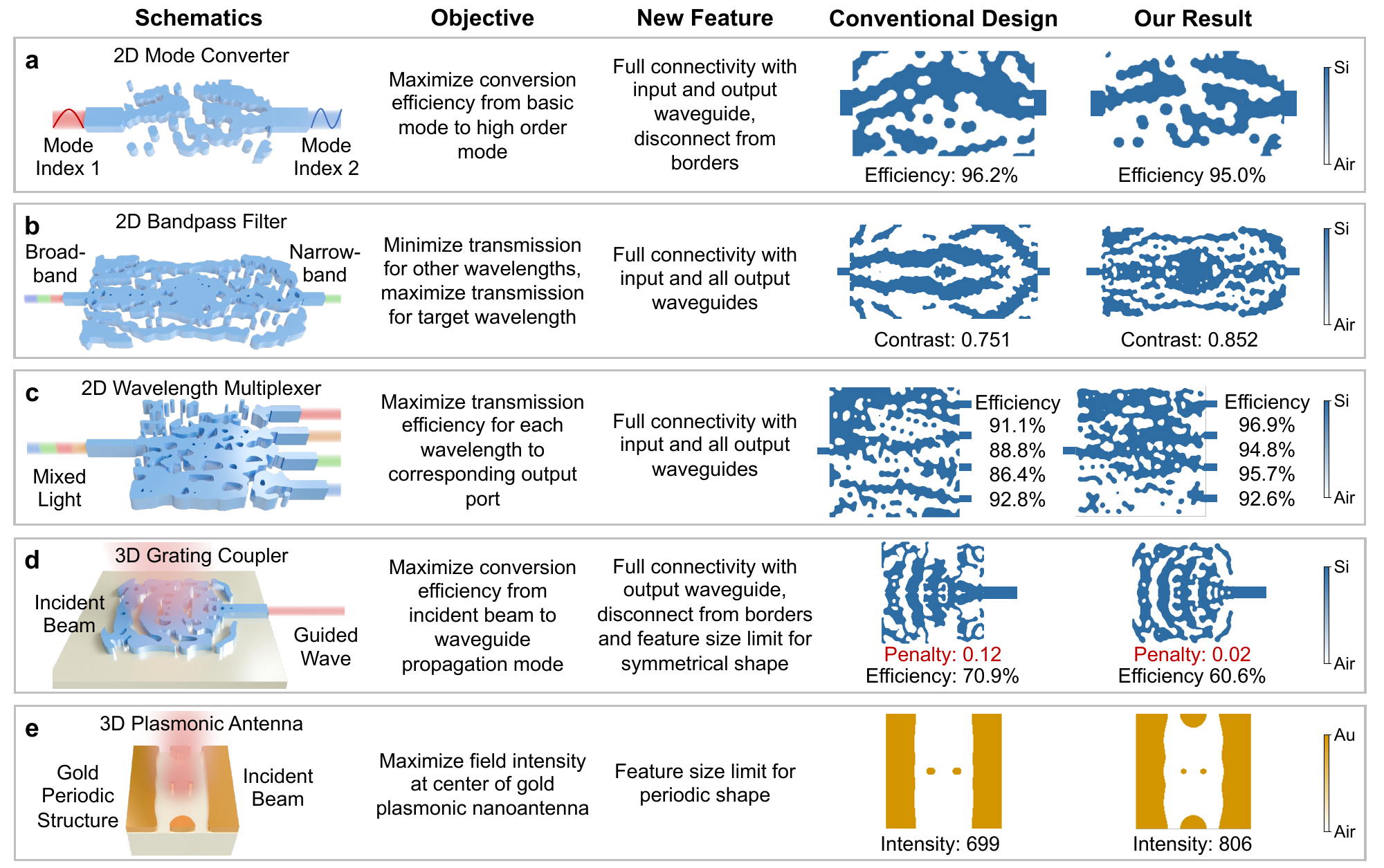}
\caption{Application of the Neuroshaper framework to various nanophotonic devices. a) 2D mode converter design demonstrating efficient conversion between different waveguide modes. The optimized structure maintains smooth connection with both input and output waveguides and is disconnected from the boundary boarders. b) 2D bandpass filter showing improved contrast ratio compared to conventional approaches while maintaining structural connectivity across all waveguide interfaces. c) 2D wavelength multiplexer design exhibiting superior transmission efficiencies compared to conventional designs while ensuring robust waveguide connectivity. d) 3D grating coupler optimization incorporating multiple constraints: connectivity with the output waveguide, boundary disconnection, and symmetry-preserving feature size limits. The conventional design violates feature size limits. e) 3D plasmonic nanoantenna optimization with periodic boundary conditions and feature size constraints, demonstrating enhanced field intensity compared to conventional designs. \textcolor{red}{Simulated field distributions for these devices can be found in Supplementary Figure 3 and Supplementary Figure 4.}}
\label{fig:5}
\end{figure}

Finally, we explore the ability and efficacy of Neuroshaper to broadly apply to the design of a wide range of free space and on-chip nanophotonic devices.  A summary of device type, objective, added constraints, and result is presented in Fig.\ref{fig:5}.  Benchmark results based on conventional pixel-based adjoint-based freeform design are also presented, using conventional spatial filtering techniques to enforce constraints.  These devices use the same initial starting point as the Neuroshaper result to provide a direct comparison. \textcolor{red}{These diverse benchmarks include common nanophotonic components: a 2D mode converter, bandpass filter, wavelength multiplexer, 3D grating coupler, and 3D plasmonic nanoantenna. Each was optimized with specific objectives (e.g., efficiency, contrast, field enhancement) and relevant constraints such as feature size, connectivity, and symmetry using Neuroshaper. For direct comparison, conventional pixel-based adjoint optimization using standard filtering techniques was performed starting from identical initial conditions. Detailed parameters for each case, including materials (typically Si/SiO2 or Au), dimensions, and specific constraints, are available in the Methods Section. }

Based on these results and benchmark comparisons, we observe the following.  First, Neuroshaper-designed devices display more robust control over feature size and curvature.  This control enforces the trade-off between geometric constraint and performance and better ensures that the devices can be fabricated.  Second, for chip-based devices, the connectivity constraint enables seamless material transitions from waveguides to freeform structures, showing how non-trivial constraints can be enforced. \textcolor{red}{For instance, while the Neuroshaper-designed 3D grating coupler (Fig.~\ref{fig:5}d) exhibits slightly lower peak efficiency (60.6\%) compared to the conventional benchmark (70.9\%), this is attributed to the enforcement of stricter, manufacturability-focused constraints within our framework, including simultaneous adherence to waveguide connectivity, and boundary disconnection requirements (see Methods Section for details on the 3D Grating Coupler benchmark). The conventional design shown achieves higher theoretical efficiency but visibly violates minimum feature size limits (note the penalty values in Fig.~\ref{fig:5}d: 0.02 for our result vs 0.12 for the conventional design) even without these stricter constraints. This highlights a key capability of Neuroshaper: balancing performance optimization with robust constraint satisfaction, which is crucial for practical device fabrication.}

\begin{figure}[htb]
\centering
\includegraphics[width=0.3\textwidth]{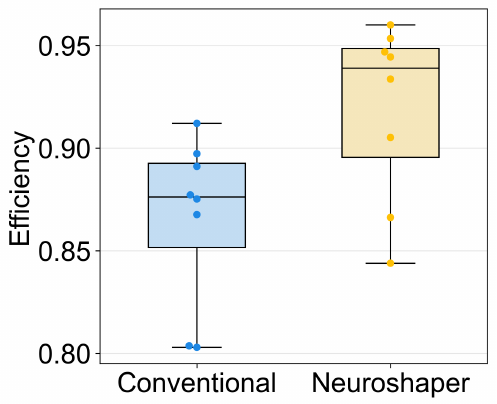}
\caption{Comparison of efficiencies for randomly initialized 2D mode converter devices. The chart displays the resulting distributions of device efficiency for eight different optimizations using Neuroshaper and conventional pixel-based methods, designed under identical conditions including initial random dielectrics, feature size and curvature constraints.}
\label{fig:6}
\end{figure}

Third, the Neuroshaper-designed devices generally have comparable or better performance than the conventional designed devices, even while featuring more total design constraints.  To more systematically probe this observation, we perform an ensemble of 2D mode converter optimizations using both Neuroshaper and conventional methods, and we fix the optimization conditions (i.e., initial random dielectrics, feature size and curvature constraints, connectivity constraints) to be the same in both cases to enable a fair conmparison.  The resulting distributions of device efficiency are shown in Figure \ref{fig:6} and show that on average the Neuroshaper-designed converters have better overall performance than conventionally designed devices.  These data suggest that Neuroshaper's ability to account for constraints in a global manner can lead to an overall improved optimization result.  

\section{Discussion}\label{sec:discussion}

In summary, we present Neuroshaper as a robust neural architecture representation framework for level set-based nanophotonic device optimization.  Our method addresses several long-standing challenges in photonic freeform inverse design.  First, it yields fully analytic, differentiable, and stable geometric representations in a manner that enables streamlining with gradient-based inverse design algorithms and accurate specification of device layouts at different scales and dielectric binarization conditions. Second, it introduces a modular approach to incorporating multiple constraints, including competing constraints.  By specifying individual constraints in the form of loss functions, constraints involving feature size, feature curvature, and connectivity can be readily imposed by appending the desired loss functions to the global loss function for network training and updating.  Third, Neuroshaper utilizes local constraints information to globally modify and evolve device layouts, providing a more robust and global framework for balancing the enforcement of constraints with device performance.  Numerical experiments indicate that Neuroshaper consistently outperforms conventional pixel-based representations in constrained freeform design tasks, and an experimental proof-of-concept demonstration indicates the viability for freeform Neuroshaper-designed devices to be experimentally translated.

\textcolor{red}{One potential limitation arises when applying Neuroshaper to very large design areas relative to the minimum feature size. The global nature of the neural representation, while advantageous for constraint handling, can sometimes lead to overly complex interdependencies across distant regions, potentially affecting optimization stability. Future work could explore architectural modifications or optimization strategies to mitigate this in large-scale problems.}  Looking forward, this work opens several promising directions for future research. We anticipate that Neuroshaper can apply to other classes of passive and active photonic devices, and more generally, to other domains of the physical sciences that leverage strong relationships between geometry and physical response.  We also anticipate that Neuroshaper can be augmented with other neural network-based algorithms \cite{YongminReview,SawyerPerspective,TingyiReview,Tahersima2019}, such as high speed deep surrogate solvers \cite{MuskensSolver,WaveyNet2022, ChenkaiICML}, generative models \cite{jiang2019free,jiang2019global,YongminMultiplexing}, and large language model interfaces, to further streamline and accelerate the computer aided design of physical infrastructure.  We see potential for neuroparameterization to encode objects other than device geometries, such as electromagnetic fields or scattering profiles, which would expand the scope of how devices and their associated attributes are modeled.  Finally, with the burgeoning growth of neurorepresentation research in the computer vision community, we anticipate a wide diversity of neural network-based representation models to be developed with practical use in the physical sciences.

\section{Methods}\label{sec:methods}
 
\subsection{Multi-Resolution Hash Encoding}
The hash encoding $\mathcal{H}\phi$ operates across $L$ resolution levels, with each level maintaining a dedicated table of feature vectors\cite{muller2022instant}. For each level $l$, input coordinates are mapped to grid vertices through:
\begin{equation}
\mathbf{v}_l = \lfloor \mathbf{x} \cdot N_l \rfloor, \quad N_l = N_\text{min} \cdot b^l
\end{equation}
where $N_l$ is the grid resolution at level $l$, $N_\text{min}$ is the size of the mose coarse grid, and $b$ is a geometric progression factor. These integer coordinates are mapped to feature indices through a spatial hash function:
\begin{equation}
h_l(\mathbf{v}) = \left(\bigoplus_{i=1}^d \pi_i v_i\right) \text{ mod } T_l
\end{equation}
where $\pi_i$ are large prime numbers ($\pi_1 = 1$, $\pi_2 = 2654435761$, $\pi_3 = 805459861$, $\pi_4 = 2097152$, $\pi_5 = 3231375$, $\pi_6 = 4278255361$) and $T_l$ is the hash table size at level $l$. This hash function effectively XORs the results of per-dimension linear congruential permutations, decorrelating the effect of dimensions on the hashed value. For better cache coherence, we set $\pi_1 = 1$, while the other prime numbers are chosen to be large and mutually coprime to minimize hash collisions in higher dimensions.

The encoding concatenates interpolated feature vectors from all levels:
\begin{equation}
\mathcal{H}_\phi(x,y) = [\text{Interp}(\phi_1[h_1(\mathbf{v})]), ..., \text{Interp}(\phi_L[h_L(\mathbf{v})])]
\end{equation}
\subsection{Hash Collision Handling}
A unique aspect of our encoding is its treatment of hash collisions. Rather than implementing explicit collision resolution through conventional methods like chaining or open addressing, our approach allows collisions to naturally occur and relies on the training process to handle them appropriately. When multiple spatial locations hash to the same feature vector, their gradients during optimization are effectively averaged. This creates an implicit form of importance sampling, where locations generating stronger gradients (typically corresponding to important features or boundaries) have a greater influence on the shared parameters.
For coarse levels where $(N_l + 1)^d < T_l$, the mapping is one-to-one and no collisions occur. At finer levels, collisions become more frequent but are distributed pseudo-randomly across space. Since each input is encoded through multiple resolution levels simultaneously, it is statistically unlikely for two distant points to collide at all levels, maintaining the encoding's ability to distinguish distinct spatial locations. To produce a  grayscale dielectric structure layout, $M_g(x,y)$, scaling and clipping operations are used:
\begin{equation}
M_g(x,y) = \text{min}\{\text{max}\{\alpha F_\theta(x,y) + \beta, 0\}, 1\}
\end{equation}
$\alpha$ and $\beta$ control the transition sharpness and bias.

\subsection{Performance Considerations}
The multi-resolution structure provides several computational advantages:

\begin{enumerate}
    \item Parameter updates are sparse and efficient, as each input affects only a small subset of feature vectors at each level.
    \item The hash table structure allows for efficient GPU implementation with predictable memory access patterns.
    \item The geometric progression of resolution levels ensures coverage of all relevant spatial scales while maintaining a manageable parameter count.
\end{enumerate}

The total number of parameters for the hash table is:
\begin{equation}
\text{Total Parameters} = L \cdot P \cdot T
\end{equation}
where $L$ is the number of levels, $P$ is the feature dimension per level, and $T$ is the hash table size per level. The hash table size $T$ provides a direct trade-off between memory usage, computational cost, and reconstruction quality.
\subsection{Gradient Computation}
The complete differentiability of our representation enables computation of important geometric properties. The gradient of the level set function is given by:
\begin{equation}
\nabla F_\theta(x,y) = \frac{\partial F_\theta}{\partial x}\mathbf{i} + \frac{\partial F_\theta}{\partial y}\mathbf{j}
\end{equation}
The mean curvature can be computed as:
\begin{equation}
\kappa = \nabla \cdot \frac{\nabla F_\theta}{\vert\nabla F_\theta\vert}
\end{equation}
These quantities are essential for implementing geometric constraints and ensuring fabrication-ready designs. The gradients are computed efficiently through automatic differentiation, propagating through both the neural network and the multi-resolution hash encoding.

\subsection{Initialization Implementation}

The initialization process in our framework consists of two distinct stages that work together to provide both controlled starting points and diverse optimization trajectories. Given an initial shape defined by a binary mask $M_0(x,y)$, we first generate an approximate level set function through a signed distance transform:
\begin{equation}
F_0(x,y) = \text{DT}(M_0(x,y))
\end{equation}
This transform computes the signed eulidean distance from each point to the nearest boundary of the initial shape, with positive values inside the shape and negative values outside. The distance transform is implemented using a fast marching method that ensures accuracy while maintaining computational efficiency.

The network parameters are then initialized using a truncated normal distribution:
\begin{equation}
\omega_i \sim \mathcal{N}(0, \sigma^2), \quad \sigma = \sqrt{\frac{2}{n_{\text{in}} + n_{\text{out}}}}
\end{equation}
where $n_{\text{in}}$ and $n_{\text{out}}$ are the input and output dimensions of each layer respectively. This initialization scheme, known as Kaiming initialization, helps maintain consistent gradient magnitudes throughout the network during the early stages of training\cite{kaiming2015delving}.

The optimization process minimizes the L1 loss between the neural network output and the initial level set function:
\begin{equation}
\mathcal{L}_{\text{init}} = \| F_{\theta}(x,y) - F_0(x,y)\|_1
\end{equation}

During this initialization phase, we employ stochastic sampling of points within the design domain to compute the loss:
\begin{equation}
\mathcal{L}_{\text{batch}} = \frac{1}{B}\sum_{i=1}^B \| F_\theta(x_i,y_i) - F_0(x_i,y_i)\|_1
\end{equation}
where B is the batch size (typically 65536 in our implementation). This stochastic approach introduces beneficial noise during initialization while maintaining computational efficiency.
The initialization process uses Adam optimization with learning rate $ = 1 \times 10^{-4}$, $\beta_1 = $ 0.9, $\beta_2 = $ 0.999

We typically run the initialization optimization for 300 steps, which we found sufficient to capture the general structure of the initial shape while maintaining enough flexibility for subsequent optimization.

For pure random initialization, we skip the distance transform step and instead initialize the network parameters directly.

\subsection{Periodicity and Symmetry Implementation}

For periodic boundaries, we apply two triangle wave mappings and then concatenating the results:
\begin{equation}
x_{\text{periodic},1} = 1 - 2\left\vert\left\{\frac{x}{p}\right\} - 0.5\right\vert
\end{equation}
\begin{equation}
x_{\text{periodic},2} = 1 - 2\left\vert\left\{\frac{x}{p} + 0.25\right\} - 0.5\right\vert
\end{equation}
where $p$ is the period length and ${\cdot}$ denotes the fractional part. The concatenated coordinates $[x_{\text{periodic},1}, x_{\text{periodic},2}]$ ensure periodicity while avoiding symmetry in the mapped domain.

For symmetry constraints, we apply a single triangle wave transformation:
\begin{equation}
x_{\text{symmetric}} = 1 - 2\left\vert\frac{x}{p} - 0.5\right\vert
\end{equation}

These mappings maintain several crucial properties. The gradient magnitude of the triangle wave remains constant except at the vertices:
\begin{equation}
\left\vert\frac{d}{dx}x_{\text{periodic},i}\right\vert = \frac{2}{p}, \quad \left\vert\frac{d}{dx}x_{\text{symmetric}}\right\vert = \frac{2}{p}
\end{equation}
This uniform gradient preserves the multi-resolution characteristics of our design representation, unlike sinusoidal mappings which introduce varying gradients:
\begin{equation}
\left\vert\frac{d}{dx}\sin\left(\frac{2\pi x}{p}\right)\right\vert = \frac{2\pi}{p}\left\vert\cos\left(\frac{2\pi x}{p}\right)\right\vert
\end{equation}

The complete coordinate transformation for a design requiring both periodicity and symmetry can be expressed as:
\begin{equation}
\mathbf{x}_{\text{transformed}} = \begin{bmatrix}
x_{\text{periodic},1} & x_{\text{periodic},2} & y_{\text{symmetric}}
\end{bmatrix}
\end{equation}
This transformation preserves feature size constraints as they operate on the raw coordinates before mapping:
\begin{equation}
L_{\text{feature}}(F_\theta(\mathbf{x}_{\text{transformed}})) = L_{\text{feature}}(F_\theta(\mathbf{x}))
\end{equation}
This ensures simultaneous satisfaction of periodicity, symmetry, and geometric requirements. The uniformity of the triangle wave gradients prevents edge artifacts while maintaining the effectiveness of our multi-resolution representation across the entire design domain.

\subsection{Feature Size and Curvature Constraint Implementation}

The feature size control implementation begins by identifying boundary points where the level set function crosses zero. For each boundary point $\mathbf{p}$, to numerically evaluate the penalty value, we employ Monte Carlo sampling with these 12 randomly distributed evaluation points along the normal direction for each of 65536 boundary points. This stochastic sampling strategy provides robust feature size control while avoiding potential artifacts that could arise from regular sampling patterns. We compute a series of randomly sampled evaluation points along the normal direction $\mathbf{n}$:
\begin{equation}
\mathbf{p}_i = \mathbf{p} + t_i d\mathbf{n}, \quad t_i \sim \mathcal{U}(-1, 1), \quad i = 0,\ldots,11
\end{equation}
where $d$ is the minimum feature size, $\mathcal{U}(-1, 1)$ denotes uniform random sampling in the interval $[-1, 1]$, and the normal direction is computed from the gradient of the level set function:
\begin{equation}
\mathbf{n} = \frac{\nabla F_\theta}{\vert\nabla F_\theta\vert}
\end{equation}
The feature size constraint combines both minimum feature size enforcement and signed distance function maintenance across all evaluation points:
\begin{equation}
\begin{aligned}
L_\text{feature}(F_\theta) = \sum_{p \in \mathcal{P}} \sum_{i=0}^{11} &[\text{ReLU}(-F_\theta(\mathbf{p}_i)\cdot\text{sign}(t_i)) \\
&+ \alpha(\vert F_\theta(\mathbf{p}_i)\vert - \vert t_i\vert d)^2]
\end{aligned}
\end{equation}
where $\mathcal{P}$ represents the set of boundary points and $\alpha$ is a weighting parameter typically set to 0.1. The first term in the summation enforces the minimum feature size by ensuring proper sign of the level set function at each evaluation point, while the second term maintains the signed distance property by encouraging the function values to match the expected distance at each evaluation point.

The curvature at boundary points is computed directly from the level set function:
\begin{equation}
\kappa = \nabla \cdot \frac{\nabla F_\theta}{\vert\nabla F_\theta\vert}
\end{equation}
The curvature penalty is formulated as a direct sum over boundary points:
\begin{equation}
L_\text{curve}(F_\theta) = \sum_{p \in \mathcal{P}} \kappa(p)
\end{equation}
The complete geometric constraint combines both feature size and curvature terms:
\begin{equation}
L_\text{geometric} = L_\text{feature} + \lambda L_\text{curve}
\end{equation}
where $\lambda$ is typically set between 0.01-0.1 depending on the specific application requirements. This random sampling approach ensures robust feature size control while avoiding systematic biases that might arise from regular sampling patterns.

\subsection{Heat Equation Formulation for Connectivity}

Our framework implements connectivity constraints through a heat equation-based approach that enables precise control over structural connectivity while maintaining compatibility with gradient-based optimization\cite{li2016structural}. The method can enforce multiple simultaneous connectivity requirements between arbitrary points or boundaries in the design space, making it particularly valuable for creating fabrication-ready metasurface designs.

The heat equation governing the connectivity constraint is formulated as:
\begin{equation} -\nabla \cdot (k\nabla u) = f \end{equation}
where $u$ represents the scalar temperature field, $k$ denotes the spatial conductivity distribution, and $f$ represents the source/sink distribution. The connectivity constraint $g$ is evaluated using a normalized p-norm of the temperature field:
\begin{equation} g = \left(\frac{\|u\|_p}{\text{thresh}}\right) - 1 \end{equation}
where $\|u\|_p = \left(\sum_i \vert u_i\vert^p\right)^{1/p}$ is the p-norm of $u$.

The framework enables sophisticated connectivity control through several key mechanisms:

First, boundary conditions can be flexibly specified by setting appropriate source ($f$) and conductivity ($k$) values at desired locations. Any point or edge in the design can be designated as a connection target through these boundary conditions. The conductivity field $k$ is directly derived from the design variables $\rho$, with high conductivity in material regions ($k_s$) and low conductivity in void regions ($k_v$).

Second, multiple connectivity requirements can be enforced simultaneously by solving multiple heat equations with different boundary conditions. For instance, to ensure connectivity between multiple waveguide interfaces or to maintain connections to multiple boundaries, separate temperature fields are computed and combined into a joint constraint. The framework automatically handles the interaction between these multiple constraints through gradient computation.

Third, the method incorporates morphological analysis to adaptively apply connectivity constraints. Before computing the heat equation solution, the current design is analyzed to identify disconnected components. If existing connections already satisfy the requirements, the constraint gradients are zeroed in those regions, focusing the optimization on areas where connectivity needs to be established or improved.

To compute the gradient $\frac{\partial g}{\partial \rho}$ for optimization, we employ the adjoint method:
\begin{equation} \frac{\partial g}{\partial \rho} = \frac{\partial g}{\partial u} \frac{\partial u}{\partial \rho} + \frac{\partial g}{\partial k} \frac{\partial k}{\partial \rho} + \frac{\partial g}{\partial f} \frac{\partial f}{\partial \rho}
\end{equation}

We introduce an adjoint variable $\lambda$ satisfying:
\begin{equation}
-\nabla \cdot (k\nabla \lambda) = \frac{\partial g}{\partial u} 
\end{equation}

The final gradient expression becomes:
\begin{equation} \frac{\partial g}{\partial \rho} = -\lambda^T \left((k_s - k_v) \nabla u \cdot \nabla u + (f_s - f_v)\right)
\end{equation}
where $k_s - k_v$ represents the difference between solid and void conductivities, and $f_s - f_v$ denotes the difference between solid and void source terms.

The numerical implementation employs finite differences for spatial derivatives and supports both direct and iterative solvers for the resulting linear systems. The heat equation is discretized using central differences, and the conductivity at material interfaces is computed using harmonic averaging to ensure numerical stability. This approach provides smooth, well-behaved gradients suitable for optimization while accurately capturing connectivity requirements.

This formulation has proven particularly effective for metasurface design, where maintaining proper connectivity is crucial for both fabrication and optical performance. The method successfully ensures that optimized structures maintain continuous paths between specified boundaries or points while allowing the topology to evolve freely in other regions. This capability is demonstrated in our results, where complex metasurface designs consistently exhibit proper connectivity to waveguide interfaces and maintain specified periodic boundary conditions.

\subsection{Resolution Conversion and Sub-pixel Averaging Details}

Starting from our continuous level set function $F_\theta(x,y)$, we implement a three-stage resolution conversion process. The first stage involves high-resolution sampling, where we sample the continuous function at an extremely high base resolution $M_b$ that significantly exceeds both geometric analysis and simulation requirements:
\begin{equation}
M_b(i,j) = \begin{cases}
1 & \text{if } F_\theta(x_i, y_j) > 0 \\
0 & \text{otherwise}
\end{cases}
\end{equation}
where $(x_i, y_j)$ represents the spatial coordinates of each high-resolution pixel.

In the second stage, we derive both the geometric analysis resolution $M_1$ and simulation resolution $M_2$ through average pooling operations from this high-fidelity binary representation:
\begin{equation}
M_k(i,j) = \frac{1}{n_k^2}\sum_{p=0}^{n_k-1}\sum_{q=0}^{n_k-1} M_b(n_ki+p,n_kj+q)
\end{equation}
Here, $k \in {1,2}$ denotes the resolution level and $n_k$ represents the downsampling factor for level $k$, with $M_1$ used for geometric constraint evaluation and $M_2$ used for electromagnetic simulation.

The third stage involves effective medium approximation for electromagnetic simulation, where the averaged values are interpreted as effective medium properties:
\begin{equation}
\epsilon_\text{eff}(i,j) = \epsilon_1 M_2(i,j) + \epsilon_2(1-M_2(i,j))
\end{equation}

In this equation, $\epsilon_\text{eff}(i,j)$ represents the effective permittivity at position $(i,j)$, while $\epsilon_1$ and $\epsilon_2$ are the permittivities of the constituent materials, and $M_2(i,j)$ represents the material fraction from average pooling.

The sub-pixel averaging approach provides several critical advantages for electromagnetic simulation. First, it offers improved accuracy by providing a more accurate representation of material boundaries than binary discretization, and this effective medium approximation also minimizes artificial resonances and scattering that can arise from staircased boundaries in binary discretization:
\begin{equation}
\text{Staircasing Error} \propto \left\vert\frac{\partial \epsilon}{\partial x}\right\vert_{\text{discrete}} - \left\vert\frac{\partial \epsilon}{\partial x}\right\vert_{\text{continuous}}
\end{equation}

Our framework also supports dynamic resolution adaptation during the optimization process. The base resolution can be adjusted throughout the optimization according to:
\begin{equation}
M_b^\text{iter}(i,j) = M_b(i\cdot s^\text{iter}, j\cdot s^\text{iter})
\end{equation}
where $s^\text{iter}$ is a scaling factor that evolves throughout the optimization:
\begin{equation}
s^\text{iter} = s_0 \cdot \alpha^{\lfloor \text{iter}/N \rfloor}
\end{equation}
This approach enables rapid exploration of the design space in early iterations using lower resolutions, followed by progressive refinement of geometric details as optimization converges, ultimately leading to high-resolution optimization for fabrication-ready designs.

\subsection{Hyperparameters of Optimization Pipeline for Nonlocal Metasurface}\label{sec:nonlocal}

\textcolor{red}{The optimization was performed using PyTorch (v2.2) on a single NVIDIA RTX A6000 GPU. Each optimization step, including constraint evaluation and backpropagation through the Neuroshaper network, typically completed in under 2 seconds, representing minimal overhead compared to the FDTD simulation time, which takes roughly 2 mins.} The hash encoding is implemented with a hierarchical structure consisting of $L=10$ levels, each encoding features of dimensionality $P=4$. The base resolution begins at $N_\text{min}=4$ and scales geometrically between consecutive levels by a factor of 1.5. Each level maintains an independent hash table of size $2^{10}$ entries. The complete encoding transforms 2D spatial coordinates into a high-dimensional feature space through parallel lookup operations at multiple resolutions.

The neural network architecture comprises a sequence of fully connected layers with dimensions [128, 64, 64], employing LeakyReLU activation functions with $\alpha=0.1$ between layers. All linear transformations are implemented without bias terms. The network maps from the hash-encoded feature space to a scalar level set value, maintaining end-to-end differentiability throughout the optimization process. The algorithm identifies boundary points through iterative optimization of $N=65536$ sample points. The optimization process alternates between gradient descent on the level set function and point resampling. At each iteration, points are categorized based on their squared SDF values, with the update strategy retaining 50\% of the best-performing points, incorporating 25\% from regular grid sampling, and introducing 25\% new random samples. Feature size control is implemented through our gradient-based sampling around boundary points. The maximum SDF value is constrained to 0.5. The feature size is considered 0.2 normalized units, for 100 nm feature size in a 500 nm device. We set $\lambda=0.1$ for the curvature penalty

The nonlocal metasurface simulation employs a uniform grid with 60 points per unit cell. The physical dimensions correspond to a period of 0.5 $\mathrm{\mu}$m in both directions, with the structure thickness set to 0.12 $\mathrm{\mu}$m. Material properties are defined by refractive indices $n_\text{Si}=3.75$ for silicon and $n_\text{SiO2}=1.45$ for the substrate. The simulation domain is terminated with perfectly matched layers (PML) of thickness 1.0 $\mathrm{\mu}$m, and excitation is provided at normal incidence ($\theta=0^\circ$) with the $H_z$ field component monitored for optimization. The framework maintains two distinct resolution levels: a high-resolution representation of $240\times240$ pixels for geometric constraint evaluation, and a lower resolution $60\times60$ grid for FDTD simulation. Conversion between resolutions is accomplished through area averaging, maintaining differentiability throughout the optimization process. We did not change the resolution over the optimization process as in the current size it already runs sufficiently fast.

The optimization employs the Adam algorithm with initial learning rates of $\eta=0.001$ for optimization. Batch sizes are set to 65,536 for geometric constraint evaluation and derived from the simulation resolution for physics optimization. The implementation maintains pure gradient-based updates without additional regularization terms.

\subsection{Experimental setup}\label{sec:setup}

For fabrication of the device, a 120 nm thick layer of hydrogenated amorphous silicon (a-Si:H) is deposited on a fused silica substrate of 500 \textmu m using plasma-enhanced chemical vapor deposition (PECVD, PlasmaTherm CCP-Dep) at a chamber temperature of 200 \textdegree C. Next, Surpass 3000 is spun on the wafer to promote adhesion of the e-beam resist 3\% hydrogen silsesquioxane (HSQ) and the latter is then spun and baked. Next, a charging dissipating solution (Electra 92) is spun. The sample is then exposed using e-beam lithography and developed in the 25\% TMAH developer, while the Electra 92 layer is solved in water. The metasurface pattern is then transferred to the a-Si:H film through dry etching (Oxford III-V Etcher) using a mixture of hydrogen bromide (HBr) and chlorine (Cl\textsubscript{2}) gases. The etch quality is controlled by adjusting the chamber pressure and forward power of the RF and ICP generators in the inductively-coupled plasma reactive-ion etching (ICP-RIE) system.

\begin{figure}[htb]
\centering
\includegraphics[width=0.7\textwidth]{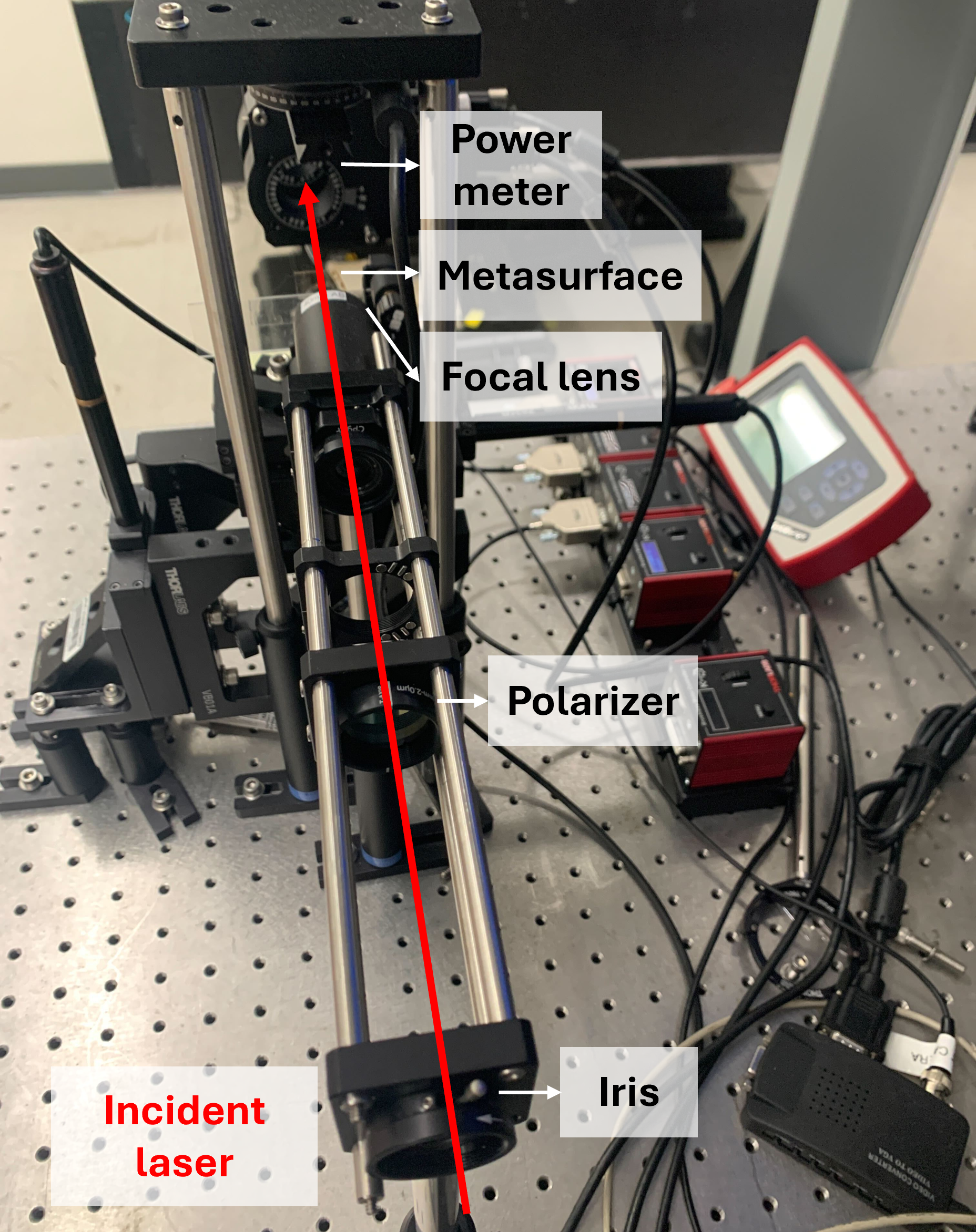}
\caption{Photograph of the experimental transmission spectra collection setup. The image shows the arrangement of optical components, including a supercontinuum white light laser, tunable bandpass filter (iris), polarizer, focusing lens, sample stage, and power meter, used for characterizing the fabricated metasurfaces.}
\label{fig:7}
\end{figure}

The experimental setup for measuring the transmission spectra of the nonlocal metasurfaces is depicted in Figure \ref{fig:7}. A collimated beam from a supercontinuum white light laser (NKT Photonics, SuperK EXTREME EXW-12) was used as the light source, with a tunable bandpass filter (NKT Photonics, SuperK LLTF) applied at the output to select specific wavelengths within the target range of 750 nm to 1000 nm. The incident beam was linearly polarized along the designed polarization of the metasurface using a polarizer and then focused through a 40 mm focal length lens to shrink the beam spot to match the dimension of the metasurface. The sample was mounted on an x-y translational stage to ensure precise alignment of the beam with individual devices. The transmitted power was measured using a power meter (Thorlabs PM100D) placed behind the sample along the optical axis. The experiment involved two sequential measurements: the transmitted power through the bare substrate was first recorded as a reference, then the beam was positioned through the metasurface, where the transmission spectrum of the metasurface is calculated as the ratio of the transmitted power through the metasurface to the reference measurement. This setup provides a robust and accurate method to characterize the transmission performance of the metasurface across a broad wavelength range.

\subsection{Details for Representative Nanophotonic Problems}\label{sec:benchmark}
Here we provide detailed setup information for the five benchmark problems used to demonstrate Neuroshaper's capabilities across different application domains. The key parameters for the 2D problems are summarized in Table~\ref{tab:benchmark_details_2d}, and for the 3D problems in Table~\ref{tab:benchmark_details_3d}.

\begin{table*}[htbp]
\small 
\centering
\caption{Detailed setup information for the 2D benchmark nanophotonic problems. ``Eval'' refers to initial evaluation and ``Ens'' to ensemble comparison.}
\label{tab:benchmark_details_2d}
\begin{tabular}{@{}llll@{}}
\toprule
Parameter & Mode Converter & Bandpass Filter & Wavelength Mux \\
\midrule
\multicolumn{4}{@{}l}{\textit{General Design Parameters}} \\
Wavelength(s) & $1.0 \, \mu\text{m}$ & $0.8-1.2 \, \mu\text{m}$ & $1.27, 1.29, 1.31, 1.33 \, \mu\text{m}$ \\
Design Region Size & $5 \times 3 \, \mu\text{m}^2$ (Eval) & $10 \times 5 \, \mu\text{m}^2$ & $4.5 \times 4.5 \, \mu\text{m}^2$ \\
& $5 \times 2 \, \mu\text{m}^2$ (Ens) & & \\
Resolution & $10 \, \text{nm}$ & $15 \, \text{nm}$ & $15 \, \text{nm}$ \\
Min. Feature Size & $240 \, \text{nm}$ & $240 \, \text{nm}$ & $200 \, \text{nm}$ \\
Symmetry & None & None & None \\
Performance Metric & Mode conv. eff. & Trans. contrast & Avg. trans. eff. \\
Field Visualization & $E_z$ & $H_z$ & $E_z$ \\
\midrule
\multicolumn{4}{@{}l}{\textit{Material Properties (Relative Permittivity, $\epsilon_r$)}} \\
Background $\epsilon_r$ & $1.0$ & $1.0$ & $1.0$ \\
Waveguide Core $\epsilon_r$ & $2.75^2$ & $2.0^2$ & $3.49^2$ \\
\midrule
\multicolumn{4}{@{}l}{\textit{Waveguide/Port Parameters}} \\
Input Wg. Width & $0.7 \, \mu\text{m}$ & $0.4 \, \mu\text{m}$ & $0.3 \, \mu\text{m}$ \\
Output Wg. Width & Same as input & $0.4 \, \mu\text{m}$ & $0.3 \, \mu\text{m}$ \\
Output Wg. Spacing & -- & -- & $0.8 \, \mu\text{m}$ \\
Input Mode & Fund. TE & Fund. TM & Fund. TE \\
Output Mode & 2nd order TE & Fund. TM & Fund. TE \\
\midrule
\multicolumn{4}{@{}l}{\textit{Other Device-Specific Parameters}} \\
Bandwidth & -- & $100 \, \text{nm}$ & -- \\
\bottomrule
\end{tabular}
\end{table*}

\begin{table*}[htbp]
\small 
\centering
\caption{Detailed setup information for the 3D benchmark nanophotonic problems. ``J\&C'' refers to Johnson and Christy data for gold permittivity.}
\label{tab:benchmark_details_3d}
\begin{tabular}{@{}lll@{}}
\toprule
Parameter & 3D Grating Coupler & 3D Plasmonic Nanoantenna \\
\midrule
\multicolumn{3}{@{}l}{\textit{General Design Parameters}} \\
Wavelength & $1.55 \, \mu\text{m}$ & $910 \, \text{nm}$ \\
Design Region Footprint & $4 \times 4 \, \mu\text{m}^2$ & $500 \times 500 \, \text{nm}^2$ \\
Design Layer Material & Si & Au \\
Design Layer Thickness & $220 \, \text{nm}$ (Si) & $50 \, \text{nm}$ (Au) \\
Resolution & $20 \, \text{nm}$ & $3 \, \text{nm}$ \\
Min. Feature Size & $160 \, \text{nm}$ & $48 \, \text{nm}$ \\
Symmetry & Mirror (xz-plane) & Mirror (xz, yz-planes) \\
Performance Metric & Coupling eff. & Max. field enhancement \\
\midrule
\multicolumn{3}{@{}l}{\textit{Material Properties (Relative Permittivity, $\epsilon_r$)}} \\
Background $\epsilon_r$ & $1.0$ & $1.0$ \\
Design Layer $\epsilon_r$ & $3.48^2$ (Si) & From J\&C (Au) \\
Substrate Material $\epsilon_r$ & $1.44^2$ (Oxide) & $1.44^2$ (SiO$_2$) \\
Substrate Thickness & $1.6 \, \mu\text{m}$ & $728 \, \text{nm}$ \\
\midrule
\multicolumn{3}{@{}l}{\textit{Device-Specific Waveguide/Optical Parameters}} \\
Input Waveguide Width & $0.5 \, \mu\text{m}$ (Si wg.) & -- \\
Output Mode & Fiber mode & -- \\
Spot Diameter & $2.5 \, \mu\text{m}$ & -- \\
Beam Tilt Angle & $10^\circ$ & -- \\
\bottomrule
\end{tabular}
\end{table*}

For each benchmark, the design region is discretized according to specified resolution, and grid resolution satisfies minimum 20 pixels per wavelength in all materials. Except for periodic dimensions, the boundary conditions are perfectly matched layers (PML). For conventional optimization, feature size constraints are enforced using erosion-dilation penalty with conic filter, and the topology optimization is performed using Adam optimizer with learning rate 0.01. A specific beta parameter for projection is ramped from 1 to 30 over optimization, which projects the parameters using the following relation.

\begin{equation}
    f(x ; \beta)=\frac{\tanh \left(\frac{\beta}{2}\right)+\tanh \left(\beta\left(x-\frac{1}{2}\right)\right)}{2 \tanh \left(\frac{\beta}{2}\right)}
\end{equation}

For comparison of the two methods, we reduced the y design region from 3 \textmu m to 2 \textmu m and removed the constraints regarding connectivity to enable a fair comparison.

\backmatter

\bmhead{Data Availability}
The data that support the findings of this study are available from the corresponding author upon reasonable request.

\bmhead{Code Availability}
The custom code and scripts used for generation and analysis of data in this study are available from the corresponding author upon reasonable request. Information on software versions includes Python (v3.10), PyTorch (v2.2), Tidy3D (v2.7). Specific variables and parameters are detailed in the Methods Section.

\bmhead{Acknowledgments}
This work was supported by the Samsung Global Outreach Program and the National Science Foundation under Award Number 2103301. CM and SA were supported by the Stanford Graduate Fellowship. Fabrication was performed in part at the Stanford Nanofabrication Facility (SNF) and the Stanford Nano Shared Facilities (SNSF), which are supported by the National Science Foundation as part of the National Nanotechnology Coordinated Infrastructure under award ECCS-1542152. 

\bmhead{Author Contributions}
T.D., Y.S., Y.Z. and J.F. developed the idea, designed experiments, and wrote the manuscript. T.D. designed and implemented the core algorithm, performed optical simulations, processed the results and prepared figures 1-6. Y.S., C.M. and S.A. performed the sample preparation. Y.W. built up the optical system and performed the optical measurement. All authors have read and approved the final manuscript.

\bmhead{Competing Interests}
The authors declare no competing interests.

\bibliography{bibliography}

\end{document}